\documentclass[aoas,MSNbibl,nameyear,dvips]{arximspdf}
\usepackage{graphicx}

\doi{10.1214/11-AOAS525}
\volume{6}
\issue{2}
\pubyear{2012}
\firstpage{542}
\lastpage{560}

\makeatletter

\newcommand{\thmu}{m_{\mu}}
\newcommand{\GG}{\mathcal{G}}
\newcommand{\thb}{\bolds{\theta}}
\newcommand{\bth}{\bolds{\theta}}
\newcommand{\bbeta}{\bolds{\beta}}
\newcommand{\eb}{\mathbf{e}}
\newcommand{\Yb}{\mathbf{Y}}
\newcommand{\Bb}{\mathbf{B}}
\newcommand{\Wb}{\mathbf{W}}
\newcommand{\bO}{\bolds{\Omega}}
\newcommand{\mbfs}{\bolds}
\newcommand{\eqref}[1]{(\ref{#1})}
\renewcommand{\citep}[1]{\citeauthor{#1} \citeyear{#1}}
\makeatother

\begin{document}
\begin{frontmatter}

\title{Modeling dependent gene expression}
\runtitle{Modeling dependent gene expression}

\begin{aug}
\author[A]{\fnms{Donatello}~\snm{Telesca}\corref{}\ead[label=e1]{donatello.telesca@gmail.com}},
\author[B]{\fnms{Peter}~\snm{M\"{u}ller}},
\author[C]{\fnms{Giovanni}~\snm{Parmigiani}}
\and
\author[D]{\fnms{Ralph~S.}~\snm{Freedman}}
\runauthor{Telesca, M\"{U}Ller, Parmigiani and Freedman}
\affiliation{UCLA School of Public Health, University of Texas M.D.
Anderson Cancer Center,
Dana Farber Cancer Institute and Harvard School of Public Health and
University of Texas M.D. Anderson Cancer Center}
\address[A]{D. Telesca\\
Department of Biostatistics\\
UCLA School of Public Health\\
Los Angeles, California 90095--1772\\
USA\\
\printead{e1}} 
\address[B]{P. M\"{u}ller\\
University of Texas Austin\\
Department of Mathematics\hspace*{42.9pt}\\
Austin, Texas 78712\\
USA}
\address[C]{G. Parmigiani\\
Department of Biostatistics\\
Harvard School of Public Health\\
Boston, Massachusetts 02115\\
USA}
\address[D]{R. S. Freedman\\
University of Texas\\
M.D. Anderson Cancer Center\\
Department of Gynecologic Oncology\\
Houston, Texas 7030\\
USA}
\end{aug}

\received{\smonth{2} \syear{2010}}
\revised{\smonth{10} \syear{2011}}

%
\begin{abstract}
In this paper we propose a Bayesian approach for inference
about dependence of high throughput gene expression. Our goals are to
use prior knowledge about pathways to anchor inference about dependence
among genes; to account for this dependence while making inferences
about differences in mean expression across phenotypes; and to explore
differences in the dependence itself across phenotypes. Useful features
of the proposed approach are a model-based parsimonious representation
of expression as an ordinal outcome, a novel and flexible representation
of prior information on the nature of dependencies, and the use of a
coherent probability model over both the structure and strength of the
dependencies of interest. We evaluate our approach through simulations
and in the analysis of data on expression of genes in the Complement and
Coagulation Cascade pathway in ovarian cancer.
\end{abstract}

%

\end{frontmatter}

\section{Introduction}\label{sec1}

Inferring patterns of dependence from high throughput
geneomic data poses significant challenges. Statistically, the problem
is one of learning about dependence structures in high dimension, with
relatively low signal. A~promising direction for strengthening this
inference is the explicit consideration of information from known
``pathways''---biochemical processes described in terms of a series of
relationships among genes and their products.

In this paper we take this perspective, and propose a Bayesian approach
to achieve three related goals in the context of gene expression
analysis: to use prior knowledge about pathways to anchor inference
about dependence among genes; to account for this dependence while
making inferences about differences in mean expression
across phenotypes; and to explore differences in the dependence itself across
phenotypes. The proposed model builds on the POE model
(\citep{Parmigiani:2002}) and integrates inference about probability
of differential expression with inference about dependence between genes
through the formulation of a coherent probability model. Our proposed
inferences are local in the sense that the model is centered around a
specific pathway.
Formally, variable selection is used to remove and add structure
relative to the centering pathway.
This is in contrast to approaches aimed at learning dependence
structures de novo from expression data, without guidance by a prior
pathway structure.

Some of the existing approaches for probabilistic modeling of dependence
structures attempt to explore the space of all possible graphical
models, often restricted to Directed Acyclic Graphs (DAGs) or Bayesian
networks (BN) (\citep{Lauritzen:1996}) and decomposable models
(\citep{Dawid:1993}). A~comprehensive review of statistical
methodology for network data is provid\-ed in \citet{Kolaczyk09}. Recent
literature includes the application of BN and dynamic BN to microarray
data (\citep{Murphy:1999}, \citep{Fried:Linal:2000}), with
applications and extensions of this methodology reported in
\citet{Ong:Glasner:2002} and \citet{Beal:Falciani:2005}, among
others. Although appealing, these techniques have computational and
methodological limitations related to modeling conditional
independence under the ``large~$p$, small~$n$'' paradigm and the difficult
specification of consistent prior models across dimensions
(\citep{Dobra:2004}). Other authors (\citep{Scott:Carvalho:2008},
\citep{Jones:2005}) have reported difficulties with the performance of
standard trans-dimensional MCMC methods (\citep{GiudiciGreen:1999}) in
the exploration of the model space, and suggested alternative stochastic
search schemes. For a decision theoretic perspective on graphical
model selection see \citet{Sebastiani05}.

To overcome these problems, we focus on variations of a baseline model
that represents known dependence structures.
The centering anchors
the model space to a prior path diagram elicited from sets of
molecular interactions derived by previous experiments.

Our idea is similar to the modeling approaches described in
\citet{Wei:Li:2007} and \citet{Wei:Li:2008}, who introduced
conditional independence between genes, via a Markov random field
defined over binary hidden states of differential expression.
These authors propose to consider a fixed Markov random field mirroring
exactly the
topology of a prior pathway and ignoring the directionality of the
edges. The construction of dependence patterns based on hidden
Markov random fields had also been previously explored by
\citet{Broet06} in the analysis of CGH microarrays.
We contrast the approach of Wei and Li (\citeyear{Wei:Li:2007,Wei:Li:2008}) in three
fundamental ways. First, we
provide an alternative interpretation of the connections encoded
into a prior pathway. We develop a prior model for the dependence
structure, that is, based on the
reciprocal graphs
(\citep{Koster:1996}).
This class of graphical models takes account of the
directionality of the edges included in the pathway and allows for
the Markovian characterization of
cycles, which often arise
in biological depictions of genetic interactions. Also,
recognizing that a known pathway is often summarizing results
obtained under different experimental conditions, we allow for
significant deviations from the prior dependence structure. This extension
requires explicit consideration of a~model determination strategy,
but enables inference on the model parameters as well as inference
on the dependence structure between genes. Finally, our focus is on
identifying significant interactions between genes in a prior
pathway, as opposed to identifying differentially expressed genes
in a given pathway.

The proposed methodology finds motivation in the analysis
of gene expression of Epithelial Ovarian Cancer (EOC) patients.
In this setting, the complement and coagulation cascade pathway
(Figure \ref{fig:Pathway}) represents a key study target, as disease progression is
thought to be
highly linked to inflammation and vascularization processes [\citet
{Wang:2005}].

The rest of this article is organized as follows. In Section
\ref{sec:POE} we introduce the proposed model. Section
\ref{sec:Inference} discusses estimation and inferential details
associated with the proposed model. We validate our approach with a
simulation study in Section~\ref{sec:simulation}. Section
\ref{sec:analysis} employs the model for the analysis of epithelial
ovarian cancer expression data, to derive inference about active genetic
interactions. In the example, a well-known molecular pathway provides
prior information on the dependence structure. A final section concludes
with a critical discussion of limitations and possible extensions.

\section{Dependent probability of expression}
\label{sec:POE}

In Section~\ref{sec:Dependence}, we discuss graphical models and
notation, and in Section \ref{subsec:POE}, we review the POE (Probability of
Expression) model \citet{Parmigiani:2002}, which defines biological
events via latent three-way indicators
of relevant biological states.
The original POE model assumes independence across genes,
conditional on hyperparameters. We
extend the original model by formalizing
more complex relationships among variables via a cascade of
conditional dependences, guided by a predefined 
interaction map.
The predefined interaction map is formally represented as a graph.
In Section \ref{sec:PG} we introduce a prior probability model on this graph.

\subsection{Representing dependence through graphical models}
\label{sec:Dependence}

Networks of relationships among expression levels can be represented as
graphs that describe how genes influence each other [for an example in
ovarian cancer see \citet{Wang:2005}]. More formally, a graph is often
characterized as an algebraic structure $\mathcal{G}=\{V,E\}$, composed
of a set of nodes $V$, in our case genes, and a~set of edges
$E \subseteq\{ (v_{i_1}, v_{i_2}), v_i \in V\} \cup
\{ \{v_{i_1}, v_{i_2}\}, v_i \in V\}$.
Here $(v_i,v_j)$ denotes a directed edge from $v_i$ to $v_j$, and
$\{v_i,v_j\}$ denotes an undirected edge.
A graph $\GG$ defines the
Markov properties of a statistical model in a graphical fashion, via the
specification of a set of conditional dependencies.

Biochemical networks often include the presence of cycles and feedback
relationships. This requires some care when trying to characterize a
coherent probabilistic model that accurately portrays prior biochemical
knowledge. For this purpose, we focus on a class of graphical models
known as reciprocal graphs [\citet{Koster:1996}].
Reciprocal graph are defined as a natural generalization of other
well-known classes, including directed acyclic graphs (DAG) and Markov random
fields, among others.
Reciprocal graphs are defined through the coherent probabilistic
interpretation of directed $a(\rightarrow b)$, undirected $(a\relbar
b)$ and reciprocal edges $(a\rightleftarrows b)$.
Here, for simplicity, we
consider a subset of the reciprocal graph family excluding
undirected edges.
The restriction to only directed edges will later be important to
facilitate the mapping of $\GG$ to a simultaneous equation model.

%
\begin{figure}[b]

\includegraphics{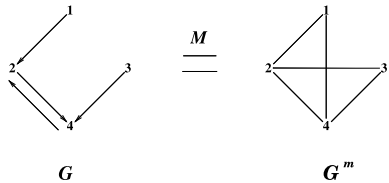}

\caption{(Example) moralization of a reciprocal graph.} \label{fig:ReciprocalMoral}
\end{figure}

The proposed model and inference is based on the directed graph
$\GG$. But sometimes it is of interest to describe the implied
conditional independence structure, that is, the Markov properties.
When desired,
the Markov properties of our model are defined in terms of an undirected
graph $\mathcal{G}^m = \{V, E^m\}$ elicited via moralization
[\citet{Koster:1996}, \citet{Lauritzen:1996}] of a graph $\GG$. In
substance, the moralization procedure consists in adding an undirected
edge between parents of a common child and replacing the remaining
directed edges with undirected ones. In $\mathcal{G}^m$, standard
Markov field properties hold, in the sense that two genes $i$ and $j$
are disconnected when they are conditionally independent, given the rest
of the genes [\citet{Besag:1974}]. For example, consider the
reciprocal graph $\mathcal{G}$ represented in Figure \ref
{fig:ReciprocalMoral}. The class of Markov equivalent models spanned by
$\mathcal{G}$ may be represented with the moral (undirected) graph
$\mathcal{G}^m$, for which the following Markov property holds:
$1\perp
2\vert 3,4$, that is, $P(1,3\vert 2,4) = P(1\vert 2,4)P(3\vert 2,4)$. The
correspondence between~$\mathcal{G}$ and~$\mathcal{G}^m$ is not 1-to-1
as $\mathcal{G}^m$ could arise from the moralization of an entire class
of Markov equivalent reciprocal graphs. Further details about
moralization in reciprocal graphs and covariance equivalence are
discussed in \citet{Koster:1996} and \citet{Spirtes:1998}. Here, our
inference will be based on $\GG$ only, and the directionality will be
based on prior knowledge. The undirected graph $\GG^m$ provides a
convenient summary of the conditional independence structure if desired.

\subsection{Dependent gene expression and hidden systems of
simultaneous equations}
\label{subsec:POE}

Following \citet{Parmigiani:2002}, we consider data in the form of an
$(p\times n)$ expression matrix $\mathbf{Y}$, with the generic element
$y_{ij}$ denoting the observed gene expression for gene $i$ in sample
$j$, $i=1,\ldots,p$ and $j=1,\ldots n$. We introduce latent variables
$e_{ij} \in\{-1,0,1\}$ indexing three possible expression categories
for each entry in $\mathbf{Y}$.
For example, if $\mathbf{Y}$ represents ratios of expression level relative
to a normal reference, they can be interpreted as
high, normal and low. Given~$e_{ij}$, for
each gene $i$ and each sample $j$ we assume a mixture
parameterized with $\theta=(\alpha_j, \mu_i, \kappa_i^-, \kappa_i^+)$
as follows:
%
%
\begin{equation}\label{eq:poe}
p\bigl(y_{ij} - (\alpha_j + \mu_i)|e_{ij}\bigr) = f_{ij}(y_{ij})\qquad\mbox{with }
\cases{
f_{-1i}= U( -\kappa_i^{-}, 0 ),\vspace*{2pt}\cr
f_{0i} = N( 0, \sigma^2_i ),\vspace*{2pt}\cr
f_{1i} = U( 0, \kappa_i^+ ).}
\end{equation}

In words, we assume that the observed expressions arise from a mixture
of a Gaussian distribution and two uniform distributions designed to
capture a broad range of departures relative to the Gaussian. The
interpretation of the Gaussian component varies depending on the
experimental design and sampling scheme, and can be trained in a
supervised way if data are available (\citep{garr:parm:2004}). When the
technology used for measuring expression has an internal reference, as
in Section~\ref{sec:analysis}, the high (low) class can be interpreted as over-
(under) expression compared to the reference.
The upcoming definition of a dependence structure will focus on
the latent~$e_{ij}$ and define dependence at the level of
these indicators. In other words, the proposed model could be
characterized as a boolean network on the latent~$e_{ij}$.

In (\ref{eq:poe}), $\alpha_j$ is a
sample-specific effect, included to adjust for systematic variation
across samples; $\mu_i$ is a gene-specific effect, modeling the overall
abundance of each gene, and $\kappa_i^-$ and $\kappa_i^+$ parameterize
the limits of variation in the tails. Finally, $\sigma^2_i$ is the
variance of the normal component of the distribution of gene $i$. We
follow \citet{Parmigiani:2002} in defining a~conditionally conjugate
prior for $\mu_i$, $\sigma^2_i$ and $\kappa_i^-$ and $\kappa_i^+$. Let
$\GG a(a,b)$ denote a~Gamma distribution with expectation $a/b$:
\begin{eqnarray*}
\label{eq:prior}
p(\mu_i | \thmu, \tau_\mu) &=&
N(\thmu, \tau_\mu), \qquad p(1/\sigma^{2}_i
| \gamma_\sigma, \lambda_\sigma) =
\mathcal{G}a(\gamma_\sigma, \lambda_\sigma), \\
 p(1/\kappa^-_i | \gamma_\kappa^-, \lambda_\kappa^- )&=&
\mathcal{G}a(\gamma_\kappa^-,\lambda_\kappa^-), \qquad
p(1/\kappa^+_i | \gamma_\kappa^+, \lambda_\kappa^+ )=
\mathcal{G}a(\gamma_\kappa^+,\lambda_\kappa^+);
\end{eqnarray*}
where $\operatorname{min}(\kappa^+_i,\kappa^-_i) > \kappa_0 \sigma_i$ and $\kappa
_0 =
5$. The restriction on $\kappa^-_i$ and $\kappa^+_i$ ensures that the
gene-specific mixture distribution has heavier tails than its normal
component, preserving interpretability of the three-way latent
classes. For the sample-specific effect $\alpha_j$, we impose an
identifiability constraint $\alpha_j\sim N(0,\tau^2_\alpha)$ with
$\sum_{j=0}^{n}\alpha_t = 0$.

Specifying a prior model for $e_{ij}$, we deviate from
\citet{Parmigiani:2002}, defining the model in terms of latent normal variables
[\citet{AlbertChib:1993}]. For each gene and sample we introduce a
latent Gaussian variable~$z_{ij}$, and define
%
%
\begin{equation}\label{eq:indicator}
e_{ij} = \cases{
1,& \quad $\mbox{if } z_{ij} > 1  \mbox{ high
expression},$\vspace*{2pt}\cr
0,&\quad $\mbox{if } -1 < z_{ij} \leq1  \mbox{ normal expression},$\vspace*{2pt}\cr
-1,&\quad\mbox{if } $z_{ij} \leq- 1 \mbox{ low expression},$
}
\end{equation}
where the distribution of $z_{ij}$ is defined by the following
simultaneous equations model (SEM):
%
%
\begin{equation}
\label{eq:SEM}
z_{ij}= m_{ij} + \sum_{k\neq i} \beta_{ik} (z_{kj} - m_{kj}) +
\varepsilon
_{ij},\qquad  i=1,\ldots,p, j=1,\ldots,n,
\end{equation}
with $\varepsilon_{ij}\sim N(0,s^2_i)$.
Let $\mathbf{Z}_j = (z_{1j},\ldots,z_{pj})'$ denote the $p$-dimensional
vector of latent probit scores associated with sample $j$. Also, let
$\mathbf{B}$ be the \mbox{$(p\times p)$} matrix whose diagonal elements are unity
and whose off-diagonal $(i,k)$ components are $-\beta_{ik}$. Provided
$\Bb$ is nonsingular,
the process above defines a proper joint probability density function
[\citet{Besag:1974}].
More
precisely, defining the marginal precision matrix $\mathbf{H}_z =
\operatorname{diag}(1/s_1, \ldots, 1/s_p)$ and $\bO= \Bb^\prime H_z \Bb$, we have
%
%
\begin{equation}
\label{eq:Pz}
P(\mathbf{Z}_j\vert\mathbf{m}_j, \bO) = \frac{
|\bO|^{1/2}}{(2\pi)^{1/2} } \exp\biggl\{-\frac{1}{2} (\mathbf{Z}_j -
{\bf
m}_j)^\prime\bO(\mathbf{Z}_j - \mathbf{m}_j)\biggr\},
\end{equation}
where $\mathbf{m}_j =(m_{1j},\ldots,m_{pj})^\prime$.

If $\mathbf{e}_j = (e_{1j},\ldots,e_{pj})^\prime$, the
implied probabilities for the indicators $e_{ij}$ are
%
%
\begin{equation}
\label{eq:Pe}
P(\mathbf{e}_j \vert\mathbf{m}_j, \bO) = \int_{A_{pj}}\cdots\int_{A_{1j}}
P(\mathbf{Z}_j \vert\mathbf{m}_j, \bO) \,d\mathbf{Z}_j,
\end{equation}
where $A_{ij}$ is the interval $(-\infty, -1]$ if $e_{ij} = -1$,
$(-1,1]$ if $e_{ij}=0$ and $(1,\infty)$ if $e_{ij}=1$.
We use notation $\pi^+_{ij} = p(z_{ij}>1 \vert y)$, $\pi^-_{ij} =
p(z_{ij}< -1 \vert y)$ and $p^\star_{ij} = \pi^+_{ij}-\pi^-_{ij}$.


In the context of this SEM, we propose to use a reciprocal graph, $\GG=
\{V,E\}$, to describe a dependence structure among the three-way
indicators~$e_{ij}$ that reflects a priori knowledge about a
pathway. Relationships between genes are captured via a set of
conditional independences over the joint distribution of the classes
$\mathbf{e}_j = (e_{ij}, i=1,\ldots,p)$.
This is implemented by structuring the matrix $\Bb$ so that the
off-diagonal element $(i,k)$ is null ($\beta_{ik} = 0$), if and only if
the edge $k \rightarrow i$ is not in $\{E\}$ [\citet{Spirtes:1998}].
The resulting\vadjust{\goodbreak} concentration matrix $\bO= \Bb^T H_z \mathbf{B}$ will
have zero off-diagonal elements $(\omega_{ik}=0)$ structured compatibly
with the Markov properties encoded in the moral graph $\GG^m=\{V,
E^m\}$ [\citet{Koster:1996}].
In summary, we use the SEM to define a probability model that
matches the conditional independence structure given by $\GG$.
The coefficients $\Bb$ of the SEM index a~family
of probability models that adhere to a given independence structure~$\GG$,
including an interpretation of the edge directions.

For each gene and sample, the mean $m_{ij}$ may be modeled as a
linear function ($m_{ij} = \mathbf{x}^\prime_j b_i$) of, say, a design
vector $\mathbf{x}_j$. This allows for comparisons across groups. For
example, if $\mathbf{x}_j = 1$ and $-1$ for samples under two different
biologic conditions, then the posterior distribution for $b_i$
formalizes inference on the differential expression of gene $i$ under
the two conditions, adjusting for the dependence among the genes.

Finally,
the autoregressive scheme in (\ref{eq:SEM}) implicitly assumes that
genetic interactions are invariant across all the cross-sample
biological variation represented in the study. Relaxing this assumption
is important and can be achieved by including an interaction term
relating the covariate or phenotype information in $\mathbf{x}_t$ with the
neighboring probit scores $\mathbf{z_{kj}}$ in \eqref{eq:SEM}.

In summary, we assume a mixture model for the observed gene
expressions~$y_{ij}$. The noisy data $y_{ij}$ is reduced to latent
trinary indicators which are used to define the dependence structure.
Because of the nonlinear shrinkage induced by the mixture
model, the $y_{ij}$ do not come from a
multivariate normal, and the patterns of dependence could be more
complex.

\subsection{Priors over graphical structures and dependence
parameters}
\label{sec:PG}
We define a prior probability model for the dependence
structure $\GG$. In words, the prior is based on a pathway diagram that
summarizes substantive prior information about the pathway of
interest. We interpret the pathway as a reciprocal graph $\GG_0 = \{V,
E_0\}$ (see example in Section \ref{sec:Dependence}). The prior on $\GG$ is defined on
the set of all graphs that can be obtained by deleting edges from
$\GG_0$. More formally, we define the model space generated by $\GG_0$
as $M(\GG_0) = \{\GG= (V,E): E\subset E_0\}$. If~$E_0$ comprises a
total number of $K$ edges, then $M(\GG_0)$ includes $D=2^K$ possible
models.

The definition of the the prior $p(\GG)$ can be seen as stating joint
probabilities for the multiple hypothesis testing problem implicitly
defined by inclusion versus exclusion of all possible edges. Following
the standard Bayesian variable selection scheme
[\citet{George:McCulloch:1993}, \citet{Brown:Vannucci:1998},
\citet{Dobra:2004}], we can consider edge inclusions as exchangeable
Bernoulli trials with common inclusion probability $\varphi$.
If
$k_{\GG}$ is the number of edges included in $\GG$, it follows that
$P(\GG\vert\varphi) =
\varphi^{k_\GG}(1-\varphi)^{K-k_\GG}$. When the inclusion
probability $\varphi$ comes from the Beta family
($\varphi\sim\mathcal{B}(a_\varphi,b_\varphi)$),
\citet{Scott:Berger:2006} and \citet{Scott:Carv:2009} show that this class
of prior model probabilities yield a strong control over the number of
``false'' edges included in $\GG$. The associated marginal prior on~$\GG$
becomes $p(\GG) = \Gamma(\kappa_\GG+ a_\varphi)\Gamma(K +
b_\varphi-
\kappa_\GG)/\Gamma(K + b_\varphi+ \kappa_\GG)$.\vadjust{\goodbreak}

A key feature of the proposed prior is the restriction to subsets of
$\GG_0$. Inference under the proposed model populates existing pathways
with probabilistic information associated with a biological system at a
temporal cross section of its dynamic. The restriction to $M(\GG_0)$ is
important to keep MCMC posterior simulation across the model space
practicable. For global searches, without restriction to a focused set
of models, trans--dimensional MCMC becomes impracticable. Local focus
does not preclude some extensions beyond $M(\GG_0)$ to facilitate
discovery of previously unknown interactions. For example, consider an
arbitrary graph $\GG$, without restriction to $M(\GG_0)$, and let
$m_\GG$ denote the number of deleted \textit{and} added edges
relative to
$\GG_0$. One could replace $k_\GG$ in the prior by $m_\GG$ and allow
for graphs beyond $\GG_0$. Little would change in the proposed
inference. But centering on models close to~$\GG_0$ is important. See
also related comments in Section \ref{sec:discussion}.

Our model is completed defining priors over the nonzero parameters
$\beta_{ij}\sim N(0, \sigma^2_\beta)$ $(i,j = 1,\ldots,p)$. This
defines a
conjugate prior for the normal mod\-el~\eqref{eq:SEM}. This formulation
is derived as a natural characterization of the SEM in (\ref{eq:SEM}).

We recognize that assuming full exchangeability over the edges
does not make active use of potential prior information on inter-gene
relationships,
possibly available through public data-bases like KEGG or Gene Ontology.
We note, however, that fine scale prior information on individual
interactions is easily included in the proposed inferential framework defining
partially exchangeable or independent Bernoulli trials
with interaction-specific inclusion probabilities, say, $\varphi_{ij}$.
If desired, the model can be extended with
$\operatorname{Beta}(a_{ij}, c_{ij})$ for $\varphi_{ij}$,
with hyperparameters chosen to reflect interaction summaries perhaps
elicited through available tools like the \texttt{R} package
\texttt{GOSim}. These elicitation processes are, however, still the
subject of active research [\citet{Frohlich2007},
\citet{Mistry2008}]. We therefore limit our analysis to purely
structural priors.

\section{Estimation and inference}
\label{sec:Inference}

\subsection{Model determination via RJ-MCMC}
\label{sec:rjmcmc} Let $\thb$ denote all population parameters and
unknown quantities directly associated with the sampling model introduced
in (\ref{eq:poe}). We implement posterior inference for ($\thb, \Bb,
\GG$) by setting up posterior MCMC simulation. We define the current
state $x=(\thb, \Bb,\mathcal{G})$ as the complete set of unknowns
and write $\pi(dx)$ short for the target posterior distribution
$p(\thb, \Bb, \GG\vert \Yb)$.

The MCMC is defined by
the following transition probabilities: (a) Update the parameter vector
$(\thb,
\Bb)$; and (b) Update $\mathcal{G}$, ensuring that the proposed graph
$\mathcal{G}^\prime$ is in the set $M\{\mathcal{G}_0\}$. This move
usually involves changes to $\Bb$ as well.

The updates in (a) follow the usual Metropolis-within-Gibbs scheme.
We sample components of $\thb$ directly from
their conditional posterior distributions (Gibbs sampling details are
reported in the \hyperref[app]{Appendix}).
We update the matrix $\Bb$ by row via multivariate random-walk
Metropolis--Hastings transition probabilities.
Let $pa(i)$ denote the parent nodes of node $i$ in the directed graph
$\GG$.
We define the $i${th} row of $\Bb$ as $\bbeta_{i}$
and
propose a new state $\bbeta_i' | \bbeta_{pa(i)} \sim
N_{|pa(i)|}(\bbeta
_i | \bbeta_{pa(i)}; V_{\beta,i}^*)$,
where $V_{\beta,i}^* = c ( 1/s_i^2 W_i^T W_i + 1/\sigma^2_\beta
I)^{-1}$. Here
$W_i$ is an $(n\times|pa(i)|)$ design matrix including all mean
adjusted probit scores
for parents of gene $i$ and $c$ is a
Metropolis--Hastings tuning parameter. For each row, this
proposal scheme changes $\Bb$ to $\Bb'$, defining
a local approximation of a reciprocal graph by a directed
acyclical graph.
Letting $\tilde{Z}$ denote a $p\times n$ matrix of mean-adjusted
probit scores,
the proposed transition is accepted with probability
\[
R(\Bb, \Bb')=\min\biggl\{1; \frac{|\Bb'|^n}{|\Bb|^n} \operatorname{etr} \biggl[-\frac{1}{2}
\tilde
{Z}^T(\Bb' - \Bb)^T H_z(\Bb' - \Bb)\tilde{Z}\biggr] \biggr\}.
\]

Some care is needed for the updates in (b), as they involve adding or
deleting an edge
in $\mathcal{G}$, therefore changing the dimension of the parameter
space. We implement a reversible jumps MCMC (RJ) [\citet{Green:1995}]:
\begin{longlist}[(ii)]
\item[(i)] Draw an edge $(k\rightarrow i)$ at random from
$E_0$.
If in the current state $\GG$, $(k\rightarrow i) \notin E$, propose the
birth of the new edge $k\rightarrow i$. If $(k\rightarrow i) \in E$,
propose the death of $k\rightarrow i$.

\item[(ii)] If we propose the birth $k\rightarrow i$, the structural
matrix $\Bb$ gets populated with a new element $\beta_{ik}' = u$,
where $u\sim q(u)$. If we propose the death of edge
$k\rightarrow i$, we simply set $\beta_{ik}^\prime=0$.
\end{longlist}
Steps (i) and (ii) generate a candidate $x'=(\Bb',\GG')$. Let $m=$
index the move proposed in step (i), and let $m^\prime$ index the
reverse move. The acceptance probability is
[\citet{Green:1995}]
%
%
\begin{equation}
\label{eq:RJMH}
R(x,x^\prime) = \min\biggl\{1,
\frac{\pi(dx^\prime)}{\pi(dx)}
\frac{q(m' \vert x^\prime)}{q(m \vert x) q(u)}
\biggr\},
\end{equation}
where $q(m \vert x)$ is the probability of proposing move $m$ when
the chain is in state~$x$, and $q(u)$ is the density function of $u$.
In general, $R(x,x')$ might include an additional factor involving the
Jacobian of a possible (deterministic) transformation of $(x,u)$ to
define $x'$.
The described RJ involves no such transformation.
The move $m$ is generated in step (i) by a uniform draw from $E_0$,
implying $q(m \vert x) = q(m' \vert x')$. Finally, $q(u)$ is the proposal p.d.f.
The acceptance probability of a birth
$R_{b}$ is then defined as
\begin{eqnarray*}
R_{b} &=& \min\biggl\{1; \frac{p(x^\prime\vert{\mathbf{Y}})}{p(x\vert\mathbf{Y})} q(u)^{-1}\biggr\}\\
&=& \min\biggl\{1; \frac{|\Bb'|^n}{|\Bb|^n} \operatorname{etr}\biggl [-\frac{1}{2} \tilde
{Z}^T(\Bb' - \Bb)^T H_z(\Bb' - \Bb)\tilde{Z}\biggr] \frac{\varphi
}{(1-\varphi
)q(u)}\biggr\}.
\end{eqnarray*}
Even though nonsingular matrices define a dense open set in $\mathbb
{R}^p$, if the proposed element $\beta_{ik}'$ of $\Bb'$ defines a
a numerically singular matrix, $\bO$ will not be positive definite and
we reject move $m'$
setting $R_b$ to zero. Given this sampling scheme, the probability of a
deletion is simply defined as
$R_d = 1/R_b$, with the roles of $x'$ and~$x$ as currently imputed and
proposed state
reversed.

\subsection{Graphical model selection}
\label{subsec:ModelSelect}

The posterior probability $p(\GG,\Bb\vert\Yb)$ and the corresponding
MCMC posterior simulation characterize our knowledge about the pathway
in the light of the data.
Based on this posterior probability, we may be interested in
selecting a representative graph $\GG$. The posterior only summarizes
the evidence for each $\GG$. It does not yet tell us which $\GG$s we
should finally report.

This model selection problem has been discussed by different
authors. \citet{Drton:2007} discuss graphical model selection from
the frequentist perspective, under the assumption that $n \geq p+1$,
while \citet{Jones:2005} or \citet{Meinshausen:2006} describe
selection techniques for problems where the sample size $n$ is small
when compared to the number of variables $p$. From a Bayesian perspective,
\citet{Scott:Carv:2009} provide a comprehensive discussion of Objective
Bayesian model selection in
Gaussian Graphical Models.

In the context of the model described in Section \ref{sec:POE},
graphical model selection can be defined by removing elements
$(k\rightarrow i) \in E_0$ specified by the prior graph
$\mathcal{G}_0=\{V,E_0\}$. This is equivalent to the vanishing of
the structural parameters $\beta_{ik}$ in the matrix
$\Bb$, characterizing the joint distribution of latent probit scores
$\mathbf{Z}$ [\citet{Ronning:1996}]. If the edge set $E_0$ has
size $|E_0| = Q$, graphical model selection involves testing $Q$ hypothesis
\[
H^0_{q}: \beta_{(q)}=0
\quad\mbox{vs.}\quad H^1_{q}: \beta_{(q)}\neq0\qquad
\mbox{for } q=1,\ldots,Q.
\]
When testing a large number of hypotheses it
is important to address possible multiplicity problems by
controlling
some predefined error rate. A popular
choice is to control the False Discovery Rate (FDR)
[\citet{Benj:Hoc:1995}]. Several authors [\citet{Scott:Carv:2009},
\citet{Scott:Carvalho:2008}, \citet{Scott:Berger:2006}] consider the
shrinkage prior defined in Section \ref{sec:rjmcmc} and report how
including edges with inclusion probability $P(\beta_{ik} \neq0) > 0.5$
(median model) yields strong control over the number of false
positives.


\section{Simulation study}
\label{sec:simulation}

We validate and illustrate the proposed method with a simulation
study with $p=50$ genes from $n=30$ samples. We define~$\mathbf{Y}$
as the $(p \times n)$ matrix of simulated mRNA intensities and
consider a~balanced design where 15 columns of
$\mathbf{Y}$ are from ``normal'' samples and 15 columns of~$\mathbf{Y}$ are associated
with ``tumor'' samples. Thus,
$\mathbf{x}_{ij} = (1,0)^\prime$ if~$y_{ij}$ is a~normal sample and
$\mathbf{x}_{ij} = (1,1)^\prime$ if~$y_{ij}$ is a~tumor sample.\vadjust{\goodbreak}

We generate simulated data $\Yb$ as follows. Given a set of latent
scores $\Wb\sim\mathcal{MN}(\mathbf{0},\mbfs{\Sigma}_z,\mathbf{I}_T)$,
where $\bO_z = \mbfs{\Sigma}_z^{-1}$ encodes a known conditional
dependence structure, and covariate effects $\mathbf{b}_i\sim
N_2(\mathbf{m}_i,\bolds{\sigma}^2_b I_2)$, we define $z_{ij} = w_{ij}
+ \mathbf{x}_{ij}^\prime\mathbf{b}_i$. We then generate the intensity
matrix $\mathbf{Y}$ from a three-way mixture of Gaussian distributions:
%
\begin{eqnarray} \label{eq:Simulation}
y_{ij} \vert z_{ij} \leq-1 &\sim& N(-4, 2^2),\nonumber\\
y_{ij} \vert z_{ij} > 3 &\sim& N(4, 2^2),\\
y_{ij} \vert-1 < z_{ij} \leq3 &\sim& N(0, 1).\nonumber
\end{eqnarray}
The precision matrix $\bO_z$ is defined as follows.
First we obtain the $p\times p$ matrix $\Bb$ by defining
$\gamma_{ij} =_d Ga(2, 1)$, $c_{ij}=\{-1, 1\}$ with $P(c_{ij}=1) =
0.5$ and~$\delta_0$ a Dirac mass at $0$, so that $\mathbf{B}_{ii} = 1$
(for $i=1,\ldots, p$), and the off-diagonal elements $\mathbf{B}_{ij} =
\pi_0 \delta_0 + (1-\pi_0) c_{ij}\gamma_{ij}$.
The simulation truth is deliberately chosen differently from the assumed
analysis model \eqref{eq:poe}.

We then generate $\bO_z$ by rescaling $\Bb^\prime\Bb$ to a correlation
matrix. The simulation model \eqref{eq:Simulation} is deliberately
different from the assumed analysis model, but still includes a
meaningful notion of true dependence structure and strength.

We use a prior Graph $\mathcal{G}_0 = \{V,E_0\}$ spanned by the set
of edges $E = E^*\cup\tilde{E}$, with $E^*$ spanning the simulation
truth of nonzero elements in $\mathbf{B}_{ij}$ (in our example $|E^*| =
50$) and $\tilde{E}$ serving as a random
mispecification set including false edges (in our example $|\tilde{E}|
= 87$).

%
\begin{figure}[t]

\includegraphics{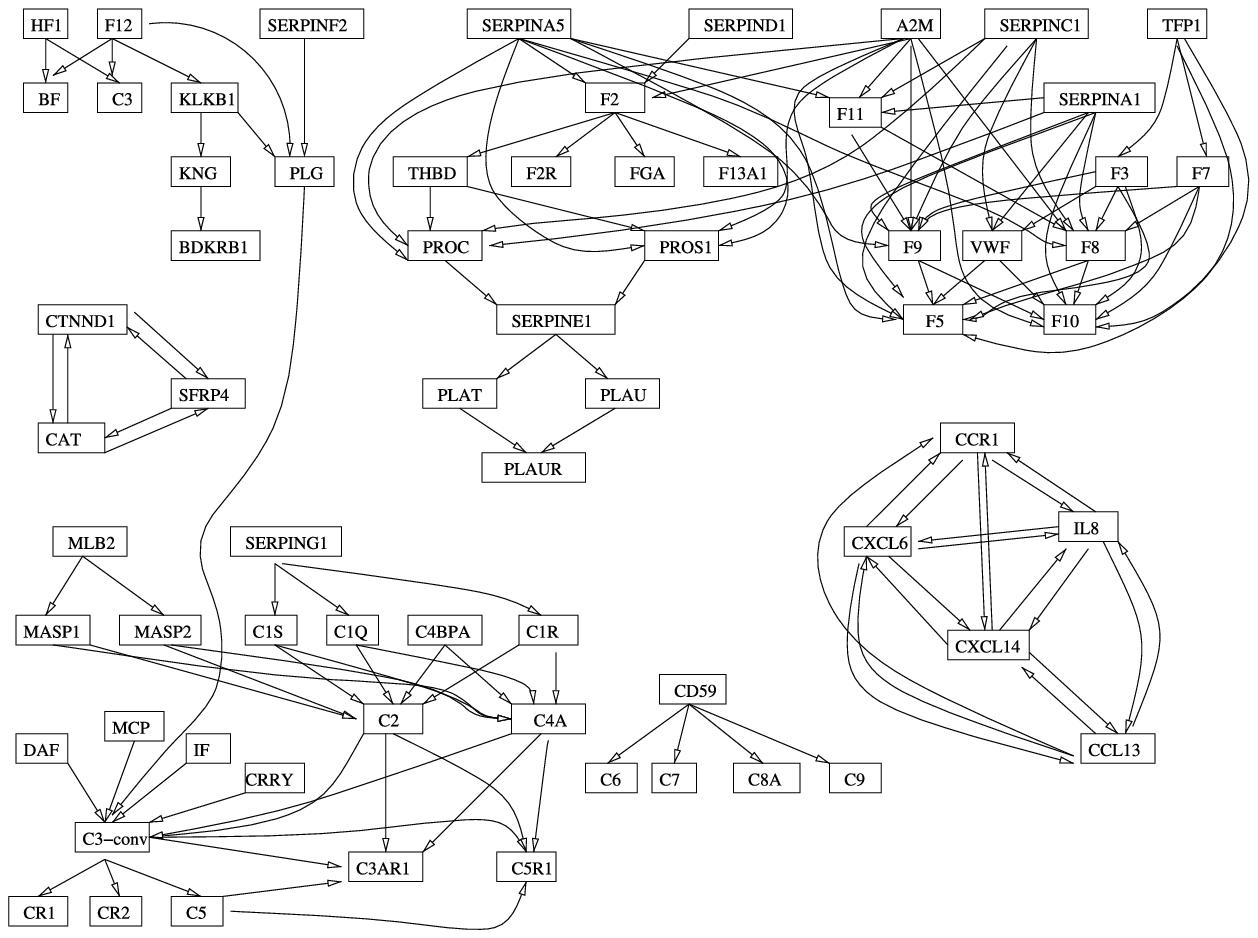}

\caption{Complement and coagulation cascades pathway [\citet{Wang:2005}].}
\label{fig:Pathway}
\end{figure}

\begin{figure}[b]

\includegraphics{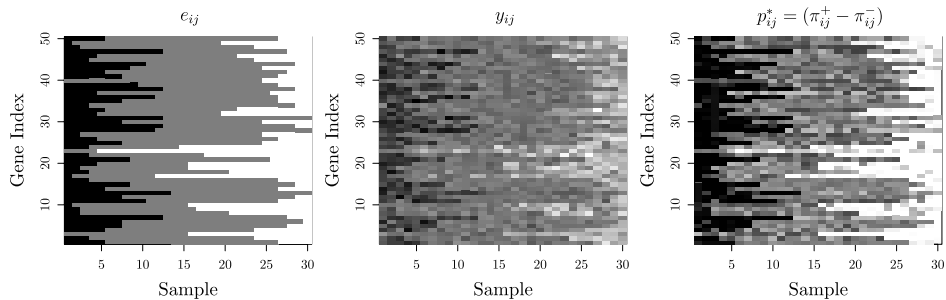}

\caption{Simulation study: (Left panel) Simulation
signal $e_{ij}$. (Central panel) Simulated mRNA abundance
$y_{ij}$. (Right panel) DepPOE estimate of $p^*_{ij}$.}
\label{fig:POEfit}
\end{figure}

In Figure \ref{fig:POEfit} we display the classification results
for the expression measurements generated under the dependence
schemes just described. We calculated posterior probabilities of over- and
under-expression from 50,000 posterior samples (thinned by 10),
obtained after conservatively discarding 50,000 iterations.
Our C$^{++}$ implementation of the algorithm, described in
Section \ref{sec:rjmcmc}, performed this simulation in about 6 hours on a standard
desktop (2.94 GHz processor).

Figure \ref{fig:POEfit} (\textit{left panel}) shows the simulation
truths as indicators ($e_{ij}$) of over- (white), normal- (grey)
and under-expression (black). The right panel reports a
unidimensional summary of the probabilities of over- or
under-expression ($p^*_{ij} = \pi^+_{ij} - \pi^-_{ij}$). The
elements $p^*_{ij}$ are defined in the $[-1,1]$ scale and may be
compared directly with the three-way indicators $e_{gt}$. We note that
the $p^*$ scale provides improved resolution over genes with signal
and recovers well the generating truth.

%
\begin{figure}

\includegraphics{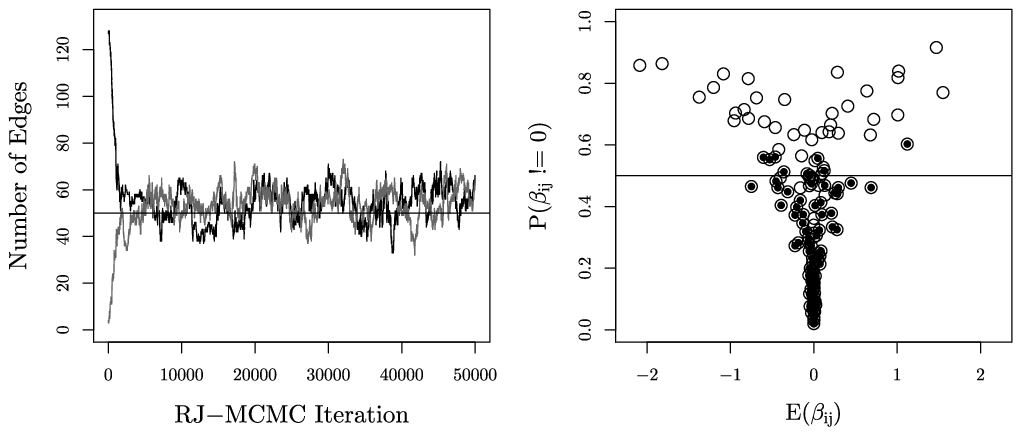}

\caption{Simulation study: (Left panel) Number of edges included in the
model by MCMC iteration, for two chains
with starting points at the two extremes of the saturation spectrum.
(Right panel) Posterior expected SEM coefficients $E(\beta_{ik} \vert
\mathbf{Y})$ vs. posterior
inclusion probabilities $P(\beta_{ik} \neq0)$. False edges are
represented with a solid circle.}
\label{fig:SimNetwork}
\end{figure}

Posterior inference includes a posterior distribution on the dependence
structure. In Figure \ref{fig:SimNetwork} (\textit{left panel}) we report
the number of edges included in the model by MCMC iteration, for two
chains starting at opposite sides of the model saturation spectrum.
Despite the size of the mispecification set $\tilde{E}$, the
trans-dimensional Markov chains converge fairly rapidly toward models
of size comparable to $|E^*| = 50$. In the same figure, marginalizing
over all possible graphs $M\{\mathcal{G}_0\}$, we report the posterior
expected SEM coefficients $E(\beta_{ik}\vert\mathbf{Y})$ and the edge
inclusion probabilities $P(\beta_{ik} \neq0 \vert\mathbf{Y})$
(\textit{right
panel}). In this plot, we report the false edges as solid
circles. Most solid circles lie in the area below an inclusion
probability of~0.5. This shows how the adopted probability scheme not
only penalizes for model complexity, but effectively controls the number
of false discoveries, allowing for a genuine recovery of the generating
conditional dependence structure.

In our simulation experiments we found that selecting edges with posterior
inclusion probabilities greater than 0.5 tends to control the false discovery
rate at level 0.01. We compared our model to the (independence) PoE model
of \citet{Parmigiani:2002} and found that including network inference as
a new inferential
goal does not diminish the classification accuracy of under- and over-expressed
samples. For details see the Web-based supplement [\citet{Telesca:2011}].
Furthermore, comparison with standard global search algorithms based
on dynamic shrinkage of partial correlation estimates point
to substantial inferential gains associated with the proposed
methodology (see Web-based supplementary
material, Section \ref{sec:Inference}).

\section{Case study}\label{sec:analysis} Wang, Wang and Kavanagh (\citeyear{Wang:2005}) report a study of epithelial
ovarian cancer (EOC). The goal of the study is to characterize the
role of the tumor microenvironment in favoring the intra-peritoneal
spread of EOC. To this end, the investigators collected tissue
samples from patients with benign (b) and malignant (m) ovarian
pathology. Specimens were collected, among other sites, from
peritoneum adjacent to the primary tumor. RNA was co-hybridized with
reference RNA to a custom made cDNA microarray including combination
of the Research genetics RG\_HsKG\_031901 8k clone set and 9000
clones selected from RG\_Hs\_seq\_ver\_070700. A complete list of
genes is available at
\texttt{\href{http://nciarray.nci.nih.gov/gal\_files/index.shtml}{http://nciarray.nci.nih.gov/gal\_files/}
\href{http://nciarray.nci.nih.gov/gal\_files/index.shtml}{index.shtml}}, ``custom
printings.'' See the array labeled Hs\_CCDTM--17.5k--1px.

In the following discussion we focus on the comparison of 10
peritoneal samples from patients with benign ovarian pathology (bPT)
versus 14 samples from patients with malignant ovarian pathology
(mPT). The raw data was processed using BRB ArrayTool
(\texttt{\href{http://linus.nci.nih.gov/BRB-ArrayTools.html}{http://linus.nci.nih.gov/BRB-}
\href{http://linus.nci.nih.gov/BRB-ArrayTools.html}{ArrayTools.html}}). In
particular, spots with minimum intensity less than 300 in both
fluorescence channels were excluded from further analysis. See
\citet{Wang:2005} for a detailed description.

One subset of genes reported on the NIH custom microarray are 61
genes in the coagulation and complement pathway
from KEGG
(\texttt{\href{http://www.genome.ad.jp}{http://www.}
\href{http://www.genome.ad.jp}{genome.ad.jp}}),
shown in Figure \ref{fig:Pathway}.
Genes on this pathway are of interest for their role in the
inflammatory process.
The arches in the pathway are
interpreted as prior judgement about (approximate) conditional
dependence (Section \ref{sec:Dependence}). However, recognizing that
the pathway represents a protein system rather than gene expression,
we allow for significant deviation from this structure, explicitly
including model determination in our analysis.

We fit the model presented in Section \ref{sec:POE} to
this set of 61 
genes. The prior set of conditional dependences between genes is
represented as a reciprocal graph in Figure \ref{fig:Pathway} and
includes a set of 148 possible edges. Reported inference is based on
50,000 MCMC samples, thinned by 10, after discarding 50,000
observation for burn--in.

\begin{figure}
\centering
\begin{tabular}{@{}c@{}}

\includegraphics{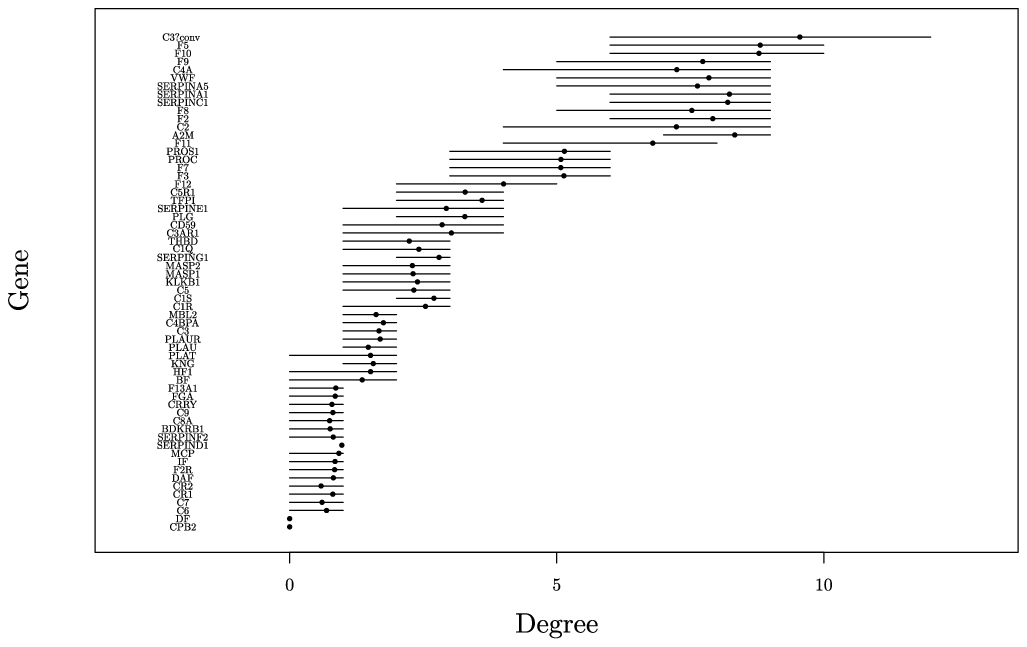}
\\
\footnotesize{(a)}\\[6pt]

\includegraphics{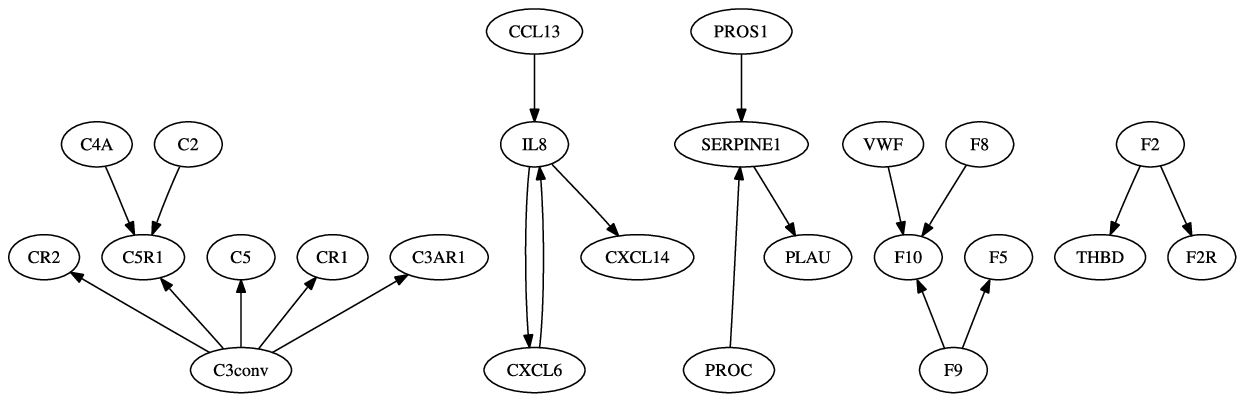}
\\
\footnotesize{(b)}
\end{tabular}
\caption{Case study. Panel (\textup{a}): Posterior mean degree and associated
95\% credible intervals by gene.
Panel (\textup{b}): Posterior pathway obtained selecting edges
with inclusion probabilities greater than 0.5 (Median model).}
\label{fig:EOC1}
\end{figure}

Recording the number of times the sampler visits a particular edge,
we calculate the posterior probability $v_{ik}=P(\beta_{ik}\vert {\mathbf{Y}})$, for each edge $(k\rightarrow i)$ in the prior
graph~$\mathcal{G}_0$. In Figure \ref{fig:EOC1} (Panel b) we show the set of
selected genetic
interactions when we consider edges with inclusion probabilities
greater than
$0.5$ (median model). Edge directionality is inherited from $\GG_0$
(Figure \ref{fig:Pathway}).

%

The posterior distribution on $\eb_g$ provides inference on
differential expression, appropriately adjusted for dependence.
Starting from the Complement and Coagulation Cascade pathway, we
identify a set of 24 genes exhibiting patterns of dependence in
their differential expression profiles across healthy and tumor
tissues.

To interpret our findings, we
searched the scientific literature using the Information Hyperlinked
Over Protein (IHOP) tool implemented by Hoffman
[\citet{Hoffman:2004}], available at: \href{http://www.ihop-net.org/UniPub/iHOP/}{www.ihop-net.org}.
For example, our study confirms the centrality of the peptide IL8
(Inteleukin-8) in the regulation of the chemokine (CXC and CC motifs)
genes. The protein encoded by this gene has been reported by several
authors to play an important role in the response to inflammatory
stimuli, resistance to apoptosis and tumoral angiogenesis. See
\citet{Terranova:Rice:1997} or \citet{Brat:Bellail:2005} for
comprehensive discussions on IL8 and its receptors.
One other example is the finding of dependent expression profiles
associated with the Thrombine pathway (F2\,{}$\rightarrow{}$F2R and
F2${}\rightarrow{}$THBD). This pathway plays a~central role in the
coagulation cascade and has been reported as a~potential mediator of
cellular function in the ovarian follicle [\citet{Roach:2002}].

Posterior edge inclusion probabilities allow for the calculation
of networks' summaries at the gene level, which summarize the role
played by individual genes in the prior pathway. In Figure \ref
{fig:EOC1} (Panel a),
we report the posterior distribution of the degree of the node
associated with each gene. This
quantity is simply defined as $d_i = p(|ne(i)| \vert \mathbf{Y})$, the
posterior distribution of
the number of neighbors associated with each gene.
This measure is often used in social science as a way to summarize
an individual's
centrality in a relational network [\citet{Sabidussi:66}]. From a
molecular biology
perspective, genes with a high degree may be interpreted as playing active
roles in the regulation of the pathway under study, in association with the
biological process of interest. Our analysis of the degree distribution
in Figure \ref{fig:EOC1} (Panel a) confirms what we observe in the selected
posterior set of genetic interactions (Panel b) and identifies important
active components of the complement and coagulation pathway.
For example, we confirm the central role of C3-convertase in the
promotion and progression of malignant ovarian cancer in humans, often reported
as a key activation component in mouse studies [\citet{Markiewski:09}].

\section{Discussion}
\label{sec:discussion}

We propose a probability model for the analysis of dependent gene
expression data. Dependence between genes is modeled via the explicit
consideration of prior information from pathways representing known
biochemical processes. We characterize a biochemical pathway
as a~reciprocal graph depicting a coherent set of conditional
dependence relationships between three-way classes of gene under-,
normal- and over-expression. Modeling dependence between latent
indicators of class membership
is likely to represent a more sensible approach for this kind of data,
when compared with methods that model correlations
between observables directly. Acknowledging that a known pathway
represents only prior information, we seek posterior inference for the
model parameters as well as for the pathway itself via an RJ-MCMC
scheme. We showed, through simulation studies, that our model enables
the recovery of the true dependence structure, even under a
misspecified prior pathway.

Our model of mRNA abundance relies on the Probability of Expression
(POE) Model of \citet{Parmigiani:2002}, and assumes that the
variability of
expression across tissue samples can be fully characterized by heavy
tailed mixtures of Normal and Uniform random variables. While this is a
simplification of reality, it contributes to denoising data and is
likely to provide useful summaries, allowing for the investigation of
the many aspects associated with expression data analysis, from data
normalization, to DE analysis, to the characterization of molecular
profiles. The general framework presented in this article is also
adaptable to other models of gene expression analysis.

In the construction of the dependent probability model, it is important
to acknowledge the limitations of the information provided in a
biochemical network. In fact, a pathway may not necessarily describe
relations among transcript levels, although it carries some information
about it. The proposed methodology is currently
restricted to known biochemical pathways. Nevertheless, structural restrictions
to one or more pathways of interest substantially simplifies
computational tractability.
The proposed model complexity is, in fact, only linear in the number of
genes and
interactions included in the prior graph. In our simulation example this
provided substantially higher power in the detection of meaningful
interactions, when compared to standard global search strategies.
Computational scalability of the proposed methodology could, however,
be an issue,
when considering highly saturated pathways including a large number of
genes. In these cases, methods based on simplifying
assumptions and approximate inference may indeed prove more
feasible as exploratory analytical tools
[\citet{Dobra:2004}].

Our model could be extended to
discover novel genetic interactions, by allowing the adding of new edges
between nodes in the prior graph $\GG_0$. This extension would,
however, come at a
substantial computational cost and would require a challenging
reformulation of the prior over graphs $p(\GG)$, to penalize for model
complexity and, at the same time, to favor models closer to the
structure of the prior pathway $\GG_0$. Initial progress in this
direction was reported by \citet{Braun08} and, in the context of Bayesian
Networks, by \citet{Mukherjee08} and it is the subject of active research.

In this article we model dependence between three-way variables as
dependence between latent Gaussian quantities. This probability scheme
is only a~convenient restriction on the possible shapes of dependence
characterizing
a~matrix of ordinal random variables. Extensions of our model
considering a~richer class of
dependence structures are, in principle, appealing. However, these
changes would
require a higher level of complexity and possible ad hock limitations
on the clique size
contributing to the joint distribution of the three-way
indicators.


%

\begin{appendix}
\section*{Appendix: Full conditional distributions}\label{app}

\textit{Sample-specific means $\alpha_t$}. From Section
\ref{subsec:POE} we have that
$p(\alpha_j \vert\tau^2_\alpha) \propto\exp\{-\alpha_j^2/  (2\tau
^2_\alpha)\}$. Using standard
conjugate analysis, it is easy to show that $ \alpha_j \vert \mathbf
{Y}_j, \bth_{\setminus\alpha_j} \sim N(\alpha_j^*, v_{\alpha
j}^*)I(l_j^*<\alpha_j < u_j^*)$,
where $v^*_{\alpha j} = \{\tau^{-2}_\alpha+ \sum_i \sigma^{-2}_g
I(e_{ij} = 0)\}^{-1}$,
$\alpha^*_{j} = v^*_{\alpha j} \sum_i \{ (y_{ij} - \mu_i)/\sigma^2_i
I(e_{ij} = 0)\}$, $l_j^* = \max_{\{g : e_{ij} = 1\}}(y_{ij} - \mu_i -
k_i^+)$ and $u_j^* = \min_{\{i : e_{ij} = -1\}}(y_{ij} - \mu_i + k_i^-)$.

\textit{Gene-specific means $\mu_i$}. From Section \ref
{subsec:POE} we have that
$p(\mu_i \vert m_\mu, \tau^2_\mu) \propto\exp\{-(\mu_i^2 - 2
m_\mu\mu
_i)/ (2\tau^2_\mu)\}$. Using standard
conjugate analysis, it is easy to show that $\mu_i \vert\mathbf{Y}_i,
\bth_{\setminus\mu_i} \sim N(\mu^*_i, v_i^*)I(l_i^*<\mu_i<u_i^*)$,
where $v^*_i = \{\tau_\mu^{-2} +\break \sigma_i^{-2}\sum_j I(e_{ij} = 0)\}
^{-1}$, $\mu_i^* = v_i^*\{m_\mu/\tau^2_\mu+ \sigma_i^{-2}\sum_j
(y_{ij} - \alpha_j)I(e_{ij = 0})\}$, $l^*_i =\break \max_{\{j : e_{ij} = 1\}
}(y_{ij} - \alpha_j - k_i^+)$ and $u_i^* = \min_{\{j : e_{ij} = -1\}
}(y_{ij} - \alpha_j + k_i^-)$.

\textit{Gene-specific variances $\sigma^2_i$}. We introduced
a conditionally conjugate Inverse Gamma prior
for $\sigma^2_i$ in Section \ref{subsec:POE}. For ease of notation we
define $h_i = 1/\sigma^2_i$, and $n_{0i} = \sum_j I(e_{ij} = 0)$. It is
easy to verify that $h_i \vert\mathbf{Y}_i, \bth_{\setminus h_i} \sim
\operatorname{Gamma}(a_i^*, b_i^*)I\{h_i \geq(\kappa_0/\min(k_i^-, k_i^+))^2\}$,
where $a^*_i = \gamma_\sigma+ n_{0i}/2$ and $b_i^* = \lambda_\sigma+
\sum_j I(e_{ij} = 0)(y_{ij} - \mu_i - \alpha_j)^2/2$.

\textit{Uniform bounds $k^{-}_i$ and $k^{+}_i$}. For ease of
notation we define
$\nu_{i0} = 1/k_i^-$ and $\nu_{i1} = 1/k_i^+$. We have $p(\nu_{i\ell})
\propto\nu_{i\ell}^{\gamma_k - 1} e^{-\lambda_k \nu_{g\ell}}$,
$(\ell
={0,1})$. The conditional posterior distribution of these parameters is
defined as $\nu_{i\ell} | \mathbf{Y}_i, \bth_{\setminus\nu_{i\ell}}
\sim \operatorname{Gamma}(a^*_{i\ell},\break b^*_{i\ell})\times I(S_{i\ell})$, where $a^*_{i\ell
} =
\gamma_k + \sum_j I (e_{ij} = 2\ell- 1)$, $b^*_{i\ell} = \lambda_k$
and $S_{i\ell} = \{\nu_{i\ell}:
\nu_{i\ell} \leq\min[ \min_{\{j: e_{ij} = 2\ell-1\}}(y_{ij} - \mu
_i -
\alpha_j), (\kappa_0\sigma_i)^{-1}]\}$.

\textit{Probit score precisions $1/s_i^2$}. For ease of
notation we define $h_{si} = 1/s_i^2$. Taking advantage of conditional
conjugacy with the distribution of probit scores, we define $p(h_{si} |
a_s, b_s) \propto h_{si}^{a_s-1} \exp\{-b_s h_{si}\}$. Let $\tilde
{z}_{ij} = z_{ij} - m_{ij} - \sum_{k\neq i} \beta_{ik}(z_{kj} -
m_{kj})'$. It is easy to show that the conditional posterior density of
$h_{si}$ is then Gamma with $p(h_{si} | \mathbf{Y}_i, \bth_{\setminus
h_{si}}) \propto h_{si}^{n/2 + a_s -1} \exp\{- h_{si}\sum_{j}(\tilde
{z}_{ij}^2)/2\}$.
\end{appendix}

\begin{supplement}[id=suppA]
\stitle{Convergence diagnostics and model comparisons\\}
\slink[doi,text={10.1214/11-\break AOAS525SUPP}]{10.1214/11-AOAS525SUPP} 
\slink[url]{http://lib.stat.cmu.edu/aoas/525/supplement.pdf}
\sdatatype{.pdf}
\sdescription{We provide an extended discussion of some aspects
associated with the proposed model. In particular, we compare
our results to the PoE model of \citet{Parmigiani:2002}
as well as some current methods used to infer networks.}
\end{supplement}

%

\printaddresses


\begin{thebibliography}{43}

\bibitem[\protect\citeauthoryear{Albert and Chib}{1993}]{AlbertChib:1993}
%
\begin{barticle}[mr]
\bauthor{\bsnm{Albert},~\bfnm{James~H.}\binits{J.~H.}} \AND
\bauthor{\bsnm{Chib},~\bfnm{Siddhartha}\binits{S.}}
(\byear{1993}).
\btitle{Bayesian analysis of binary and polychotomous response data}.
\bjournal{J. Amer. Statist. Assoc.}
\bvolume{88}
\bpages{669--679}.
\bid{issn={0162-1459}, mr={1224394}}
\bptok{imsref}%
\end{barticle}
%
\endbibitem

\bibitem[\protect\citeauthoryear{Beal et~al.}{2005}]{Beal:Falciani:2005}
%
\begin{barticle}[auto:STB|2012/01/18|07:48:53]
\bauthor{\bsnm{Beal},~\bfnm{M.}\binits{M.}},
\bauthor{\bsnm{Falciani},~\bfnm{F.}\binits{F.}},
\bauthor{\bsnm{Ghahramani},~\bfnm{Z.}\binits{Z.}},
\bauthor{\bsnm{Rangel},~\bfnm{C.}\binits{C.}} \AND
\bauthor{\bsnm{Wild},~\bfnm{D.}\binits{D.}}
(\byear{2005}).
\btitle{A Bayesian approach to reconstructing genetic regulatory
networks with
hidden factors}.
\bjournal{Bioinformatics}
\bvolume{21}
\bpages{349--356}.
\bptok{imsref}%
\end{barticle}
%
\endbibitem

\bibitem[\protect\citeauthoryear{Benjamini and
Hochberg}{1995}]{Benj:Hoc:1995}
%
\begin{barticle}[mr]
\bauthor{\bsnm{Benjamini},~\bfnm{Yoav}\binits{Y.}} \AND
\bauthor{\bsnm{Hochberg},~\bfnm{Yosef}\binits{Y.}}
(\byear{1995}).
\btitle{Controlling the false discovery rate: A practical and powerful approach
to multiple testing}.
\bjournal{J. Roy. Statist. Soc. Ser. B}
\bvolume{57}
\bpages{289--300}.
\bid{issn={0035-9246}, mr={1325392}}
\bptok{imsref}%
\end{barticle}
%
\endbibitem

\bibitem[\protect\citeauthoryear{Besag}{1974}]{Besag:1974}
%
\begin{barticle}[mr]
\bauthor{\bsnm{Besag},~\bfnm{Julian}\binits{J.}}
(\byear{1974}).
\btitle{Spatial interaction and the statistical analysis of lattice systems}.
\bjournal{J.~Roy. Statist. Soc. Ser. B}
\bvolume{36}
\bpages{192--236}.
\bid{issn={0035-9246}, mr={0373208}}
\bptnote{check related}%
\bptok{imsref}%
\end{barticle}
%
\endbibitem

\bibitem[\protect\citeauthoryear{Brat, Bellail and
Erwin}{2005}]{Brat:Bellail:2005}
%
\begin{barticle}[auto:STB|2012/01/18|07:48:53]
\bauthor{\bsnm{Brat},~\bfnm{D.~J.}\binits{D.~J.}},
\bauthor{\bsnm{Bellail},~\bfnm{A.~C.}\binits{A.~C.}} \AND
\bauthor{\bsnm{Erwin},~\bfnm{G.~V.~M.}\binits{G.~V.~M.}}
(\byear{2005}).
\btitle{The role of interlukin-8 and its receptors in gliomagenesis
and tumoral
angiogenesis}.
\bjournal{Neuro-Oncology}
\bvolume{7}
\bpages{122--133}.
\bptok{imsref}%
\end{barticle}
%
\endbibitem

\bibitem[\protect\citeauthoryear{Braun, Cope and Parmigiani}{2008}]{Braun08}
%
\begin{barticle}[auto:STB|2012/01/18|07:48:53]
\bauthor{\bsnm{Braun},~\bfnm{R.}\binits{R.}},
\bauthor{\bsnm{Cope},~\bfnm{L.}\binits{L.}} \AND
\bauthor{\bsnm{Parmigiani},~\bfnm{G.}\binits{G.}}
(\byear{2008}).
\btitle{Identigying differential correlation in gene/pathway combinations}.
\bjournal{BMC Bioinformatics}
\bvolume{9}
\bpages{488}.
\bptok{imsref}%
\end{barticle}
%
\endbibitem

\bibitem[\protect\citeauthoryear{Bro{\"{e}}t and Richardson}{2006}]{Broet06}
%
\begin{barticle}[pbm]
\bauthor{\bsnm{Bro{\"{e}}t},~\bfnm{Philippe}\binits{P.}} \AND
\bauthor{\bsnm{Richardson},~\bfnm{Sylvia}\binits{S.}}
(\byear{2006}).
\btitle{Detection of gene copy number changes in CGH microarrays using a
spatially correlated mixture model}.
\bjournal{Bioinformatics}
\bvolume{22}
\bpages{911--918}.
\bid{doi={10.1093/bioinformatics/btl035}, issn={1367-4803}, pii={btl035},
pmid={16455750}}
\bptok{imsref}%
\end{barticle}
%
\endbibitem

\bibitem[\protect\citeauthoryear{Brown, Vannucci and
Fearn}{1998}]{Brown:Vannucci:1998}
%
\begin{barticle}[mr]
\bauthor{\bsnm{Brown},~\bfnm{P.~J.}\binits{P.~J.}},
\bauthor{\bsnm{Vannucci},~\bfnm{M.}\binits{M.}} \AND
\bauthor{\bsnm{Fearn},~\bfnm{T.}\binits{T.}}
(\byear{1998}).
\btitle{Multivariate {B}ayesian variable selection and prediction}.
\bjournal{J. R. Stat. Soc. Ser. B Stat. Methodol.}
\bvolume{60}
\bpages{627--641}.
\bid{doi={10.1111/1467-9868.00144}, issn={1369-7412}, mr={1626005}}
\bptok{imsref}%
\end{barticle}
%
\endbibitem

\bibitem[\protect\citeauthoryear{Carvalho and Scott}{2009}]{Scott:Carv:2009}
%
\begin{barticle}[mr]
\bauthor{\bsnm{Carvalho},~\bfnm{C.~M.}\binits{C.~M.}} \AND
\bauthor{\bsnm{Scott},~\bfnm{J.~G.}\binits{J.~G.}}
(\byear{2009}).
\btitle{Objective {B}ayesian model selection in {G}aussian graphical models}.
\bjournal{Biometrika}
\bvolume{96}
\bpages{497--512}.
\bid{doi={10.1093/biomet/asp017}, issn={0006-3444}, mr={2538753}}
\bptok{imsref}%
\end{barticle}
%
\endbibitem

\bibitem[\protect\citeauthoryear{Dawid and Lauritzen}{1993}]{Dawid:1993}
%
\begin{barticle}[mr]
\bauthor{\bsnm{Dawid},~\bfnm{A.~P.}\binits{A.~P.}} \AND
\bauthor{\bsnm{Lauritzen},~\bfnm{S.~L.}\binits{S.~L.}}
(\byear{1993}).
\btitle{Hyper-{M}arkov laws in the statistical analysis of decomposable
graphical models}.
\bjournal{Ann. Statist.}
\bvolume{21}
\bpages{1272--1317}.
\bid{doi={10.1214/aos/1176349260}, issn={0090-5364}, mr={1241267}}
\bptok{imsref}%
\end{barticle}
%
\endbibitem

\bibitem[\protect\citeauthoryear{Dobra et~al.}{2004}]{Dobra:2004}
%
\begin{barticle}[mr]
\bauthor{\bsnm{Dobra},~\bfnm{Adrian}\binits{A.}},
\bauthor{\bsnm{Hans},~\bfnm{Chris}\binits{C.}},
\bauthor{\bsnm{Jones},~\bfnm{Beatrix}\binits{B.}},
\bauthor{\bsnm{Nevins},~\bfnm{Joseph~R.}\binits{J.~R.}},
\bauthor{\bsnm{Yao},~\bfnm{Guang}\binits{G.}} \AND
\bauthor{\bsnm{West},~\bfnm{Mike}\binits{M.}}
(\byear{2004}).
\btitle{Sparse graphical models for exploring gene expression data}.
\bjournal{J. Multivariate Anal.}
\bvolume{90}
\bpages{196--212}.
\bid{doi={10.1016/j.jmva.2004.02.009}, issn={0047-259X}, mr={2064941}}
\bptok{imsref}%
\end{barticle}
%
\endbibitem

\bibitem[\protect\citeauthoryear{Drton and Perlman}{2007}]{Drton:2007}
%
\begin{barticle}[mr]
\bauthor{\bsnm{Drton},~\bfnm{Mathias}\binits{M.}} \AND
\bauthor{\bsnm{Perlman},~\bfnm{Michael~D.}\binits{M.~D.}}
(\byear{2007}).
\btitle{Multiple testing and error control in {G}aussian graphical model
selection}.
\bjournal{Statist. Sci.}
\bvolume{22}
\bpages{430--449}.
\bid{doi={10.1214/088342307000000113}, issn={0883-4237}, mr={2416818}}
\bptok{imsref}%
\end{barticle}
%
\endbibitem

\bibitem[\protect\citeauthoryear{Friedman et~al.}{2000}]{Fried:Linal:2000}
%
\begin{barticle}[auto:STB|2012/01/18|07:48:53]
\bauthor{\bsnm{Friedman},~\bfnm{N.}\binits{N.}},
\bauthor{\bsnm{Linial},~\bfnm{M.}\binits{M.}},
\bauthor{\bsnm{Nachman},~\bfnm{I.}\binits{I.}} \AND
\bauthor{\bsnm{Pe`er},~\bfnm{D.}\binits{D.}}
(\byear{2000}).
\btitle{Using Bayesian networks to analyze expression data}.
\bjournal{J. Comput. Biol.}
\bvolume{7}
\bpages{601--620}.
\bptok{imsref}%
\end{barticle}
%
\endbibitem

\bibitem[\protect\citeauthoryear{Fr{\"o}hlich et~al.}{2007}]{Frohlich2007}
%
\begin{barticle}[auto:STB|2012/01/18|07:48:53]
\bauthor{\bsnm{Fr{\"o}hlich},~\bfnm{H.}\binits{H.}},
\bauthor{\bsnm{Speer},~\bfnm{N.}\binits{N.}},
\bauthor{\bsnm{Poutska},~\bfnm{A.}\binits{A.}} \AND
\bauthor{\bsnm{Beibart},~\bfnm{T.}\binits{T.}}
(\byear{2007}).
\btitle{GOSim--an R-package for computation of theoretic GO similarities
between terms and ene products}.
\bjournal{BMC Bioinformatics}
\bvolume{8}
\bpages{166}.
\bptok{imsref}%
\end{barticle}
%
\endbibitem

\bibitem[\protect\citeauthoryear{Garrett and
Parmigiani}{2004}]{garr:parm:2004}
%
\begin{barticle}[mr]
\bauthor{\bsnm{Garrett},~\bfnm{Elizabeth~S.}\binits{E.~S.}} \AND
\bauthor{\bsnm{Parmigiani},~\bfnm{Giovanni}\binits{G.}}
(\byear{2004}).
\btitle{A nested unsupervised approach to identifying novel molecular
subtypes}.
\bjournal{Bernoulli}
\bvolume{10}
\bpages{951--969}.
\bid{doi={10.3150/bj/1106314845}, issn={1350-7265}, mr={2108038}}
\bptok{imsref}%
\end{barticle}
%
\endbibitem

\bibitem[\protect\citeauthoryear{George and
McCulloch}{1993}]{George:McCulloch:1993}
%
\begin{barticle}[auto:STB|2012/01/18|07:48:53]
\bauthor{\bsnm{George},~\bfnm{E.~I.}\binits{E.~I.}} \AND
\bauthor{\bsnm{McCulloch},~\bfnm{R.~E.}\binits{R.~E.}}
(\byear{1993}).
\btitle{Variable selection via Gibbs sampling}.
\bjournal{J.~Amer. Statist. Assoc.}
\bvolume{88}
\bpages{881--889}.
\bptok{imsref}%
\end{barticle}
%
\endbibitem

\bibitem[\protect\citeauthoryear{Giudici and Green}{1999}]{GiudiciGreen:1999}
%
\begin{barticle}[mr]
\bauthor{\bsnm{Giudici},~\bfnm{Paolo}\binits{P.}} \AND
\bauthor{\bsnm{Green},~\bfnm{Peter~J.}\binits{P.~J.}}
(\byear{1999}).
\btitle{Decomposable graphical {G}aussian model determination}.
\bjournal{Biometrika}
\bvolume{86}
\bpages{785--801}.
\bid{doi={10.1093/biomet/86.4.785}, issn={0006-3444}, mr={1741977}}
\bptok{imsref}%
\end{barticle}
%
\endbibitem

\bibitem[\protect\citeauthoryear{Green}{1995}]{Green:1995}
%
\begin{barticle}[mr]
\bauthor{\bsnm{Green},~\bfnm{Peter~J.}\binits{P.~J.}}
(\byear{1995}).
\btitle{Reversible jump {M}arkov chain {M}onte {C}arlo computation and
{B}ayesian model determination}.
\bjournal{Biometrika}
\bvolume{82}
\bpages{711--732}.
\bid{doi={10.1093/biomet/82.4.711}, issn={0006-3444}, mr={1380810}}
\bptok{imsref}%
\end{barticle}
%
\endbibitem

\bibitem[\protect\citeauthoryear{Hoffman and Valencia}{2004}]{Hoffman:2004}
%
\begin{barticle}[auto:STB|2012/01/18|07:48:53]
\bauthor{\bsnm{Hoffman},~\bfnm{R.}\binits{R.}} \AND
\bauthor{\bsnm{Valencia},~\bfnm{A.}\binits{A.}}
(\byear{2004}).
\btitle{A gene network for navigating the literature}.
\bjournal{Nature Genetics}
\bvolume{36}
\bpages{664--664}.
\bptok{imsref}%
\end{barticle}
%
\endbibitem

\bibitem[\protect\citeauthoryear{Jones et~al.}{2005}]{Jones:2005}
%
\begin{barticle}[mr]
\bauthor{\bsnm{Jones},~\bfnm{Beatrix}\binits{B.}},
\bauthor{\bsnm{Carvalho},~\bfnm{Carlos}\binits{C.}},
\bauthor{\bsnm{Dobra},~\bfnm{Adrian}\binits{A.}},
\bauthor{\bsnm{Hans},~\bfnm{Chris}\binits{C.}},
\bauthor{\bsnm{Carter},~\bfnm{Chris}\binits{C.}} \AND
\bauthor{\bsnm{West},~\bfnm{Mike}\binits{M.}}
(\byear{2005}).
\btitle{Experiments in stochastic computation for high-dimensional graphical
models}.
\bjournal{Statist. Sci.}
\bvolume{20}
\bpages{388--400}.
\bid{doi={10.1214/088342305000000304}, issn={0883-4237}, mr={2210226}}
\bptok{imsref}%
\end{barticle}
%
\endbibitem

\bibitem[\protect\citeauthoryear{Kolaczyk}{2009}]{Kolaczyk09}
%
\begin{bbook}[mr]
\bauthor{\bsnm{Kolaczyk},~\bfnm{Eric~D.}\binits{E.~D.}}
(\byear{2009}).
\btitle{Statistical Analysis of Network Data: Methods and Models}.
\bpublisher{Springer}, \baddress{New York}.
\bid{doi={10.1007/978-0-387-88146-1}, mr={2724362}}
\bptok{imsref}%
\end{bbook}
%
\endbibitem

\bibitem[\protect\citeauthoryear{Koster}{1996}]{Koster:1996}
%
\begin{barticle}[mr]
\bauthor{\bsnm{Koster},~\bfnm{J.~T.~A.}\binits{J.~T.~A.}}
(\byear{1996}).
\btitle{Markov properties of nonrecursive causal models}.
\bjournal{Ann. Statist.}
\bvolume{24}
\bpages{2148--2177}.
\bid{doi={10.1214/aos/1069362315}, issn={0090-5364}, mr={1421166}}
\bptok{imsref}%
\end{barticle}
%
\endbibitem

\bibitem[\protect\citeauthoryear{Lauritzen}{1996}]{Lauritzen:1996}
%
\begin{bbook}[mr]
\bauthor{\bsnm{Lauritzen},~\bfnm{Steffen~L.}\binits{S.~L.}}
(\byear{1996}).
\btitle{Graphical Models}.
\bseries{Oxford Statistical Science Series}
\bvolume{17}.
\bpublisher{Clarendon Press}, \baddress{New York}.
\bid{mr={1419991}}
\bptok{imsref}%
\end{bbook}
%
\endbibitem

\bibitem[\protect\citeauthoryear{Markiewski and
Lambris}{2009}]{Markiewski:09}
%
\begin{barticle}[auto:STB|2012/01/18|07:48:53]
\bauthor{\bsnm{Markiewski},~\bfnm{M.~M.}\binits{M.~M.}} \AND
\bauthor{\bsnm{Lambris},~\bfnm{J.~D.}\binits{J.~D.}}
(\byear{2009}).
\btitle{Unwelcome complement}.
\bjournal{Cancer Research}
\bvolume{69}
\bpages{6367}.
\bptok{imsref}%
\end{barticle}
%
\endbibitem

\bibitem[\protect\citeauthoryear{Meinshausen and
B{\"u}hlmann}{2006}]{Meinshausen:2006}
%
\begin{barticle}[mr]
\bauthor{\bsnm{Meinshausen},~\bfnm{Nicolai}\binits{N.}} \AND
\bauthor{\bsnm{B{\"u}hlmann},~\bfnm{Peter}\binits{P.}}
(\byear{2006}).
\btitle{High-dimensional graphs and variable selection with the lasso}.
\bjournal{Ann. Statist.}
\bvolume{34}
\bpages{1436--1462}.
\bid{doi={10.1214/009053606000000281}, issn={0090-5364}, mr={2278363}}
\bptok{imsref}%
\end{barticle}
%
\endbibitem

\bibitem[\protect\citeauthoryear{Mistry and Pavlidis}{2008}]{Mistry2008}
%
\begin{barticle}[auto:STB|2012/01/18|07:48:53]
\bauthor{\bsnm{Mistry},~\bfnm{M.}\binits{M.}} \AND
\bauthor{\bsnm{Pavlidis},~\bfnm{P.}\binits{P.}}
(\byear{2008}).
\btitle{Gene ontology term overlap as a measure of gene functional similarity}.
\bjournal{BMC Bioinformatics}
\bvolume{9}
\bpages{327}.
\bptok{imsref}%
\end{barticle}
%
\endbibitem

\bibitem[\protect\citeauthoryear{Mukherjee and Speed}{2008}]{Mukherjee08}
%
\begin{barticle}[auto:STB|2012/01/18|07:48:53]
\bauthor{\bsnm{Mukherjee},~\bfnm{S.}\binits{S.}} \AND
\bauthor{\bsnm{Speed},~\bfnm{T.~P.}\binits{T.~P.}}
(\byear{2008}).
\btitle{Netrwork inference using informative priors}.
\bjournal{PNAS}
\bvolume{105}
\bpages{14133--14318}.
\bptok{imsref}%
\end{barticle}
%
\endbibitem

\bibitem[\protect\citeauthoryear{Murphy and Mian}{1999}]{Murphy:1999}
%
\begin{bmisc}[auto:STB|2012/01/18|07:48:53]
\bauthor{\bsnm{Murphy},~\bfnm{K.}\binits{K.}} \AND
\bauthor{\bsnm{Mian},~\bfnm{S.}\binits{S.}}
(\byear{1999}).
\bhowpublished{Modeling gene expression data using dynamic Bayesian
networksayesian networks.
Technical report,
Computer Science Division, Univ. California, Berkley}.
\bptok{imsref}%
\end{bmisc}
%
\endbibitem

\bibitem[\protect\citeauthoryear{Ong, Glasner and
Page}{2002}]{Ong:Glasner:2002}
%
\begin{barticle}[auto:STB|2012/01/18|07:48:53]
\bauthor{\bsnm{Ong},~\bfnm{I.}\binits{I.}},
\bauthor{\bsnm{Glasner},~\bfnm{J.}\binits{J.}} \AND
\bauthor{\bsnm{Page},~\bfnm{D.}\binits{D.}}
(\byear{2002}).
\btitle{Modelling regulatory pathways in e.coli from time series expression
profiles}.
\bjournal{Bioinformatics}
\bvolume{18}
\bpages{S241--S248}.
\bptok{imsref}%
\end{barticle}
%
\endbibitem

\bibitem[\protect\citeauthoryear{Parmigiani et~al.}{2002}]{Parmigiani:2002}
%
\begin{barticle}[mr]
\bauthor{\bsnm{Parmigiani},~\bfnm{Giovanni}\binits{G.}},
\bauthor{\bsnm{Garrett},~\bfnm{Elizabeth~S.}\binits{E.~S.}},
\bauthor{\bsnm{Anbazhagan},~\bfnm{Ramaswamy}\binits{R.}} \AND
\bauthor{\bsnm{Gabrielson},~\bfnm{Edward}\binits{E.}}
(\byear{2002}).
\btitle{A~statistical framework for expression-based molecular classification
in cancer}.
\bjournal{J. R. Stat. Soc. Ser. B Stat. Methodol.}
\bvolume{64}
\bpages{717--736}.
\bid{doi={10.1111/1467-9868.00358}, issn={1369-7412}, mr={1979385}}
\bptnote{check related}%
\bptok{imsref}%
\end{barticle}
%
\endbibitem

\bibitem[\protect\citeauthoryear{Roach et~al.}{2002}]{Roach:2002}
%
\begin{barticle}[auto:STB|2012/01/18|07:48:53]
\bauthor{\bsnm{Roach},~\bfnm{L.~E.}\binits{L.~E.}},
\bauthor{\bsnm{Petrik},~\bfnm{J.~J.}\binits{J.~J.}},
\bauthor{\bsnm{Plante},~\bfnm{L.}\binits{L.}} \AND
\bauthor{\bsnm{LaMarre},~\bfnm{J.}\binits{J.}}
(\byear{2002}).
\btitle{Thrombin generation and presence of thrombin in ovarian follicles}.
\bjournal{Biology of Reproduction}
\bvolume{66}
\bpages{1350--1358}.
\bptok{imsref}%
\end{barticle}
%
\endbibitem

\bibitem[\protect\citeauthoryear{Ronning and Kukuk}{1996}]{Ronning:1996}
%
\begin{barticle}[mr]
\bauthor{\bsnm{Ronning},~\bfnm{Gerd}\binits{G.}} \AND
\bauthor{\bsnm{Kukuk},~\bfnm{Martin}\binits{M.}}
(\byear{1996}).
\btitle{Efficient estimation of ordered probit models}.
\bjournal{J.~Amer. Statist. Assoc.}
\bvolume{91}
\bpages{1120--1129}.
\bid{issn={0162-1459}, mr={1424612}}
\bptok{imsref}%
\end{barticle}
%
\endbibitem

\bibitem[\protect\citeauthoryear{Sabidussi}{1966}]{Sabidussi:66}
%
\begin{barticle}[mr]
\bauthor{\bsnm{Sabidussi},~\bfnm{Gert}\binits{G.}}
(\byear{1966}).
\btitle{The centrality index of a graph}.
\bjournal{Psychometrika}
\bvolume{31}
\bpages{581--603}.
\bid{issn={0033-3123}, mr={0205879}}
\bptok{imsref}%
\end{barticle}
%
\endbibitem

\bibitem[\protect\citeauthoryear{Scott and Berger}{2006}]{Scott:Berger:2006}
%
\begin{barticle}[mr]
\bauthor{\bsnm{Scott},~\bfnm{James~G.}\binits{J.~G.}} \AND
\bauthor{\bsnm{Berger},~\bfnm{James~O.}\binits{J.~O.}}
(\byear{2006}).
\btitle{An exploration of aspects of {B}ayesian multiple testing}.
\bjournal{J. Statist. Plann. Inference}
\bvolume{136}
\bpages{2144--2162}.
\bid{doi={10.1016/j.jspi.2005.08.031}, issn={0378-3758}, mr={2235051}}
\bptok{imsref}%
\end{barticle}
%
\endbibitem

\bibitem[\protect\citeauthoryear{Scott and
Carvalho}{2008}]{Scott:Carvalho:2008}
%
\begin{barticle}[mr]
\bauthor{\bsnm{Scott},~\bfnm{James~G.}\binits{J.~G.}} \AND
\bauthor{\bsnm{Carvalho},~\bfnm{Carlos~M.}\binits{C.~M.}}
(\byear{2008}).
\btitle{Feature-inclusion stochastic search for {G}aussian graphical models}.
\bjournal{J. Comput. Graph. Statist.}
\bvolume{17}
\bpages{790--808}.
\bid{doi={10.1198/106186008X382683}, issn={1061-8600}, mr={2649067}}
\bptok{imsref}%
\end{barticle}
%
\endbibitem

\bibitem[\protect\citeauthoryear{Sebastiani and Ramoni}{2005}]{Sebastiani05}
%
\begin{barticle}[mr]
\bauthor{\bsnm{Sebastiani},~\bfnm{Paola}\binits{P.}} \AND
\bauthor{\bsnm{Ramoni},~\bfnm{Marco}\binits{M.}}
(\byear{2005}).
\btitle{Normative selection of {B}ayesian networks}.
\bjournal{J.~Multivariate Anal.}
\bvolume{93}
\bpages{340--357}.
\bid{doi={10.1016/j.jmva.2004.03.005}, issn={0047-259X}, mr={2162642}}
\bptok{imsref}%
\end{barticle}
%
\endbibitem

\bibitem[\protect\citeauthoryear{Spirtes et~al.}{1998}]{Spirtes:1998}
%
\begin{barticle}[auto:STB|2012/01/18|07:48:53]
\bauthor{\bsnm{Spirtes},~\bfnm{P.}\binits{P.}},
\bauthor{\bsnm{Richardson},~\bfnm{T.~S.}\binits{T.~S.}},
\bauthor{\bsnm{Meek},~\bfnm{C.}\binits{C.}},
\bauthor{\bsnm{Scheines},~\bfnm{R.}\binits{R.}} \AND
\bauthor{\bsnm{Glymour},~\bfnm{C.}\binits{C.}}
(\byear{1998}).
\btitle{Using path diagrams as a structural equation modeling tool}.
\bjournal{Sociol. Methods Res.}
\bvolume{27}
\bpages{182--225}.
\bptok{imsref}%
\end{barticle}
%
\endbibitem

\bibitem[\protect\citeauthoryear{Telesca et~al.}{2011}]{Telesca:2011}
%
\begin{bmisc}[auto:STB|2012/01/18|07:48:53]
\bauthor{\bsnm{Telesca},~\bfnm{D.}\binits{D.}},
\bauthor{\bsnm{M{\"u}ller},~\bfnm{P.}\binits{P.}},
\bauthor{\bsnm{Parmigiani},~\bfnm{G.}\binits{G.}} \AND
\bauthor{\bsnm{Freedman},~\bfnm{R.~S.}\binits{R.~S.}}
(\byear{2011}).
\bhowpublished{Supplement to ``Modeling dependent gene expression.''
DOI:\doiurl{10.1214/11-AOAS525SUPP}.}
\bptok{imsref}%
\end{bmisc}
%
\endbibitem

\bibitem[\protect\citeauthoryear{Terranova and
Rice}{1997}]{Terranova:Rice:1997}
%
\begin{barticle}[auto:STB|2012/01/18|07:48:53]
\bauthor{\bsnm{Terranova},~\bfnm{P.~F.}\binits{P.~F.}} \AND
\bauthor{\bsnm{Rice},~\bfnm{V.~M.}\binits{V.~M.}}
(\byear{1997}).
\btitle{Review: Cytokine involvement in ovarian processes}.
\bjournal{American Journal of Reproductive Immunology}
\bvolume{37}
\bpages{50--63}.
\bptok{imsref}%
\end{barticle}
%
\endbibitem

\bibitem[\protect\citeauthoryear{Wang, Wang and Kavanagh}{2005}]{Wang:2005}
%
\begin{barticle}[auto:STB|2012/01/18|07:48:53]
\bauthor{\bsnm{Wang},~\bfnm{X.}\binits{X.}},
\bauthor{\bsnm{Wang},~\bfnm{E.}\binits{E.}} \AND
\bauthor{\bsnm{Kavanagh},~\bfnm{J.}\binits{J.}}
(\byear{2005}).
\btitle{Ovarian cancer, the coagulation pathway, and inflammation}.
\bjournal{Journal of Translational Medicine}
\bvolume{3}
\bpages{25}.
\bptok{imsref}%
\end{barticle}
%
\endbibitem

\bibitem[\protect\citeauthoryear{Wei and Li}{2007}]{Wei:Li:2007}
%
\begin{barticle}[auto:STB|2012/01/18|07:48:53]
\bauthor{\bsnm{Wei},~\bfnm{Z.}\binits{Z.}} \AND
\bauthor{\bsnm{Li},~\bfnm{H.}\binits{H.}}
(\byear{2007}).
\btitle{A Markov random field model for network--based analysis of genomic
data}.
\bjournal{Bioinformatics}
\bvolume{23}
\bpages{1357--1544}.
\bptok{imsref}%
\end{barticle}
%
\endbibitem

\bibitem[\protect\citeauthoryear{Wei and Li}{2008}]{Wei:Li:2008}
%
\begin{barticle}[mr]
\bauthor{\bsnm{Wei},~\bfnm{Zhi}\binits{Z.}} \AND
\bauthor{\bsnm{Li},~\bfnm{Hongzhe}\binits{H.}}
(\byear{2008}).
\btitle{A hidden spatial-temporal {M}arkov random field model for network-based
analysis of time course gene expression data}.
\bjournal{Ann. Appl. Stat.}
\bvolume{2}
\bpages{408--429}.
\bid{doi={10.1214/07--AOAS145}, issn={1932-6157}, mr={2415609}}
\bptok{imsref}%
\end{barticle}
%
\endbibitem

\end{thebibliography}
\end{document}